\documentclass[twocolumn,10pt]{IEEEtran}
\usepackage{algorithm,algorithmic,amsbsy,amsmath,amssymb,epsfig,bbm,mathrsfs,multirow,amsthm,epstopdf,subfig}
\usepackage{color}

\newcommand{\beq}{\begin{equation}}
\newcommand{\eeq}{\end{equation}}
\newcommand{\bitm}{\begin{itemize}}
\newcommand{\ba}{\begin{array}}
\newcommand{\ea}{\end{array}}
\newcommand{\eitm}{\end{itemize}}
\newcommand{\beqn}{\begin{eqnarray}}
\newcommand{\eeqn}{\end{eqnarray}}
\newcommand{\beqno}{\begin{eqnarray*}}
\newcommand{\eeqno}{\end{eqnarray*}}
\newcommand{\bma}{\begin{displaymath}}
\newcommand{\ema}{\end{displaymath}}
\newcommand{\bnu}{\begin{enumerate}}
\newcommand{\enu}{\end{enumerate}}
\newcommand{\bce}{\begin{center}}
\newcommand{\ece}{\end{center}}
\newcommand{\btb}{\begin{tabular}}
\newcommand{\etb}{\end{tabular}}

\newtheorem{mypro}{Proposition}

\begin{document}

\title{Market Model and Optimal Pricing Scheme of Big Data and Internet of Things (IoT)}
\author{Dusit Niyato$^1$, Mohammad Abu Alsheikh$^1$, Ping Wang$^1$, Dong In Kim$^2$, and Zhu Han$^3$	\\
$^1$ School of Computer Engineering, Nanyang Technological University (NTU), Singapore	\\
$^2$ School of Information and Communication Engineering, Sungkyunkwan University (SKKU), Korea		\\
$^3$ Department of Electrical and Computer Engineering, University of Houston, Texas, USA 	\vspace{-5mm}	}

\maketitle

\begin{abstract}
Big data has been emerging as a new approach in utilizing large datasets to optimize complex system operations. Big data is fueled with Internet-of-Things (IoT) services that generate immense sensory data from numerous sensors and devices. While most current research focus of big data is on machine learning and resource management design, the economic modeling and analysis have been largely overlooked. This paper thus investigates the big data market model and optimal pricing scheme. We first study the utility of data from the data science perspective, i.e., using the machine learning methods. We then introduce the market model and develop an optimal pricing scheme afterward. The case study shows clearly the suitability of the proposed data utility functions. The numerical examples demonstrate that big data and IoT service provider can achieve the maximum profit through the proposed market model.
\end{abstract}


\section{Introduction}
\label{sec:introduction}

With an introduction of Internet of Things (IoT), a number of devices are available and/or deployed to collect data. This leads to a rapid growth of data volumes created. Together with the advancement of telecommunication, computers, storage, and algorithms, the data can be transferred, processed, stored, and analyzed efficiently. This paradigm is known as big data. In this paradigm, a large dataset is analyzed using different algorithms including machine learning to extract useful knowledge and information and to support services to businesses and customers. A number of algorithms, tools, and services have been proposed and developed to support big data management and processing~\cite{zhang2015}. In this perspective, data is a precious resource, and the concept of data-as-a-service (DaaS) has been introduced~\cite{oliveira2015}. In DaaS, data is considered as a commodity that can be accessed and processed to obtain knowledge, used in decision making. While the main stream research in big data and DaaS focuses on developing algorithms of knowledge extraction and resource management, subtle attention has been paid to an economic perspective of big data. With an emerging big data regime, market mechanisms and economic models will be crucial not only to generate a revenue and optimize resource utilization of different stockholders such as data sources, big data brokers, and service providers, but also to maximize the satisfaction of service users and knowledge consumers.

Data can be treated as a resource that can be traded in a market. With DaaS, data is acquired in an on-demand basis depending on the requirements of users and consumers. The goal of this paper is to introduce a suitable big data and IoT market model. We first introduce some background of big data analytics and data science to accentuate the importance of data as a valuable resource for knowledge extraction. We then propose candidate utility functions of data. We present a case study of classification-based machine learning algorithms to justify the concave increasing function of the utility function. Next, we introduce a market model composed of sensors, data source, service provider, and consumers (subscribers). The optimal service subscription fee is derived based on the data utility functions. The optimal pricing can be modeled as a Stackelberg game to maximize the profit of the data source. Finally, numerical results show the optimal subscription fee and optimal data price. This paper serves as an initial study to the economic analysis of big data and IoT systems.



\section{Related Work}
\label{sec:relatedwork}

Pricing is used as an economic mechanism not only for revenue generation, but also for efficient resource allocation, for example, in cloud resource~\cite{gohad2013}, wireless network~\cite{gizelis2011}, and IoT~\cite{alfagih2013}. In big data and IoT era, data becomes a resource that can be traded in a market. The authors in~\cite{pantelis2013} discussed different aspects of values of big data. They highlighted that big data involves information goods which can be sold in a market. The concept of data trading was presented, and different issues, e.g., monopoly power due to differentiation of data and information bundling, were outlined. Due to the special nature of information goods, the concept of information economics has been developed. The first issue is the method to quantify value of information. The information has its value when it is used in decision making. For example, the authors in~\cite{turgut2013} analyzed the value of information in energy-constrained intruder tracking sensor networks. The networks aim to identify and locate an intruder. If the network successfully detects the intruder, the cost that can be avoided quantifies the value of information. Based on the value of information analysis, the pricing scheme of the information can be developed. In~\cite{balasubramanian2015}, the authors developed a pricing scheme for information goods. Two information access schemes, i.e., subscription and pay-per-use, were considered, and it was shown that there are competitive prices that maximize profits of data sellers. 

One of the applications of information pricing is in cognitive radio. The authors in~\cite{duan2011} introduced a pricing model for spectrum sensing information. Secondary users obtain/buy the information so that they can opportunistically access idle licensed spectrum efficiently. However, none of the related work considered the data demand and pricing especially when the data is proposed and used to offer services. Thus, there is an immediate need to develop such a big data market model and optimal pricing scheme, which are the objectives of this paper.


\section{Big Data Analytics and Utility of Data}
\label{sec:value}

In this section, we describe the big data (data science) approach. Then, we establish the generic utility functions of data based on machine learning.

\subsection{Big Data Analytics}

\begin{figure}[h]
\begin{center}
$\begin{array}{c} 
\epsfxsize=2.5 in \epsffile{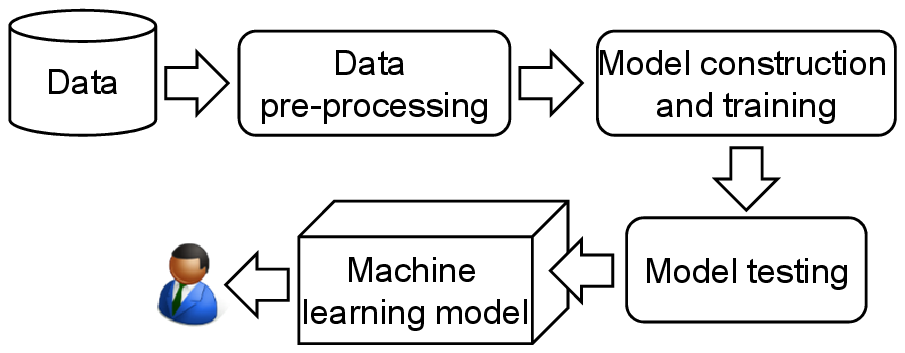} \\ 
\end{array}$
\caption{Knowledge discovery and extraction process of big data~\cite{fayyad1996}.}
\label{fig:machine_learning}
\end{center}
\end{figure}

\begin{figure*}[t]
\begin{center}
$\begin{array}{cc} 
\epsfxsize=3.4 in \epsffile{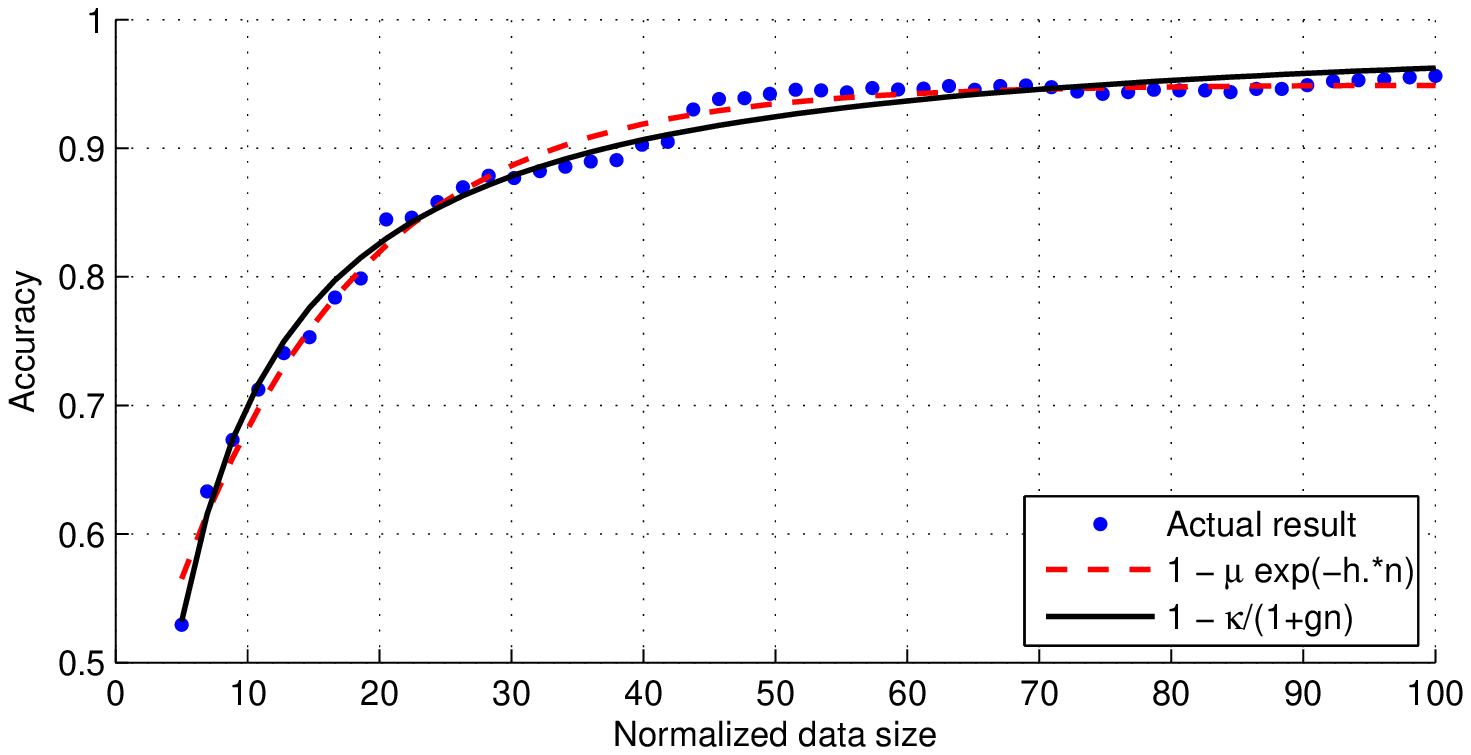} &	
\epsfxsize=3.4 in \epsffile{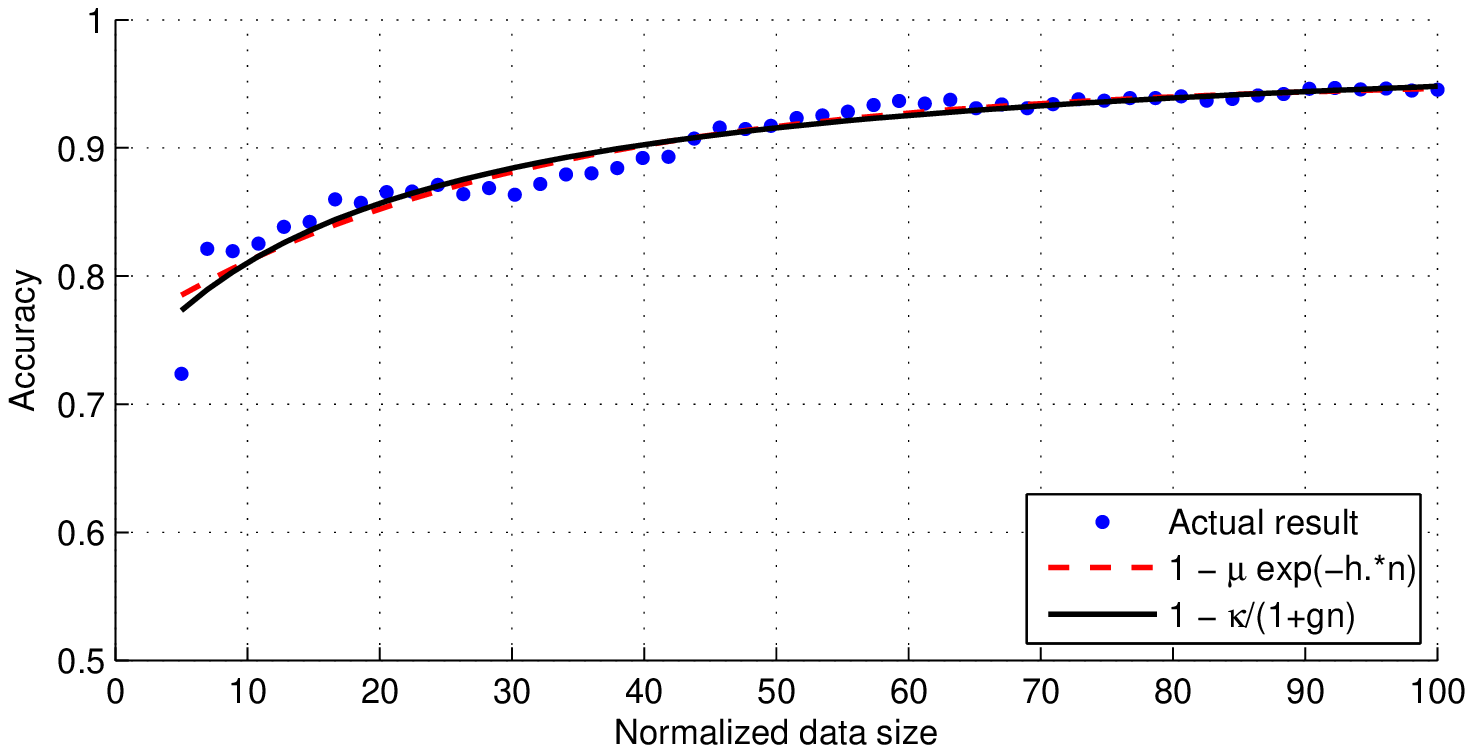}	\\ 
(a)	&	(b)		\\	
\end{array}$
\caption{Estimation of the data-service quality function $\phi(n)$ using (a)~support vector machines ($\kappa=1.109, g=0.271, \mu=0.457, \text{and } h=0.039$), and (b)~logistic regression ($\kappa=0.273, g=0.044, \mu=0.21, \text{and } h=0.017$).}
\label{fig:activity_accuracy}
\end{center}
\end{figure*}

Big data is a form of data science that aims to use a large dataset to solve a real-world problem. Big data defines the model-based approach to extract knowledge and information that can be a solution of a specific problem of complex systems. Such complex systems cannot be easily modeled mathematically or simulated. Figure~\ref{fig:machine_learning} shows a knowledge discovery and extraction process of big data analytics. From Fig.~\ref{fig:machine_learning}, raw data plays an important role in big data analytics in that it is used to construct machine learning and data mining models. From an economic perspective, in a simplest form, we define a utility function of raw data as follows:
\begin{equation}
	u = \phi( n )	,
\end{equation}
which is a mapping from the raw data with size $n$ items to utility $u$. Machine learning and data mining algorithms use raw data to train the model. Thus, the utility can be considered as the quality of the model. For example, for regression models, the utility is the accuracy to predict a real-value target variable. Similarly, for classification, the utility is the accuracy of classifying input into a discrete-value output. It is well known that machine learning algorithms achieve better accuracy when they are trained with more raw data~\cite{domingos2012few}. Thus, we make the following reasonable assumptions of the utility function:
\begin{itemize}
	\item $\phi'(n) > 0$: utility is an increasing function;
	\item $\phi''(n) \leq 0$: the function has decreasing marginal utility.
\end{itemize}

\subsection{Case Study of Classification-Based Machine Learning}

\subsubsection{Classification-Based Machine Learning}
To determine the utility function of raw data in big data analytics, we consider a case study of classification-based machine learning algorithms. An annotated training (raw) data tuple has the form $(\vec{\mathbf{x}}_{i},y_{i})\in \mathbb{R}^{R} \times \left\{1,\ldots,C\right\}$, where $\vec{\mathbf{x}}$ is a data vector of length $R$, $y$ is a tuple label from $C$ possible classes, and $i$ is an index. The training data is collected as $\mathcal{D}=\left\{ (\vec{\mathbf{x}}_{i},y_{i})|i=1,\ldots,n\right\}$, where $n$ is the size of raw data to be used in model training. Consider a classification learning model $\varPsi:\vec{\mathbf{x}}_{i}\rightarrow \hat{y}_{i}$ which is trained on the raw data. The trained model maps input data into a predicted label $\hat{y}_{i}$. An effective classifier is designed to minimize the error between $y_{i}$ and $\hat{y}_{i}$. For example, minimize the residual sum of squares, i.e., $\min\sum_{i=1}^{n}\left\Vert y_{i}-\hat{y}_{i}\right\Vert _{2}^{2}$.

\subsubsection{Estimating Utility Functions}

After the classifier model is trained with the given raw data, it is tested to determine the accuracy, e.g., using the separate sets of test data. The classification error is defined as $\epsilon_{j}$ for raw data with size $n_{j}$, where $j$ is the index of the test instance. Without loss of generality, we assume that the utility, i.e., accuracy, is $u_j = 1-\epsilon_j$. To estimate the utility function $\phi(n)$, we vary the size of raw data used in the model training. In particular, the experiment points are $(n_{1},\epsilon_{1}),\ldots, (n_j,\epsilon_j),\ldots,(n_{L},\epsilon_{L})$ for $n_j < n_{j+1}$ and $j = 1,\ldots,L-1$. These points are then used to find a set of optimal parameters of the candidate utility function $\phi(n;\mathbf{\alpha})$ by non-linear least squares, where $\mathbf{\alpha}$ contains the parameters. The non-linear least squares algorithm optimizes $\mathbf{\alpha}$ by minimizing the sum of square errors as follows:
\begin{equation}
\min_{\mathbf{\alpha}}\sum_{j=1}^{L}\Big(\epsilon_{j}- \big( 1 -   \phi(n_{j};\mathbf{\alpha}) \big) \Big)^2.
\end{equation}

We consider two candidate utility functions:
\begin{itemize}
	\item The fraction-based function is defined as follows:
		\begin{equation}
			\phi(n;\mathbf{\alpha}=[\kappa,g]) = 1 - \frac{\kappa}{1 + g n},
		\end{equation}
		for $\frac{\kappa}{1 + g n} \leq 1$, where $\kappa$ and $g$ are curve fitting parameters. These fitting parameters are adjusted such that the sum of the squared errors between the experimental and estimated points is minimized. 
	\item The exponential-based function is defined as follows:
		\begin{equation}
			\phi(n;\mathbf{\alpha}=[\mu,h]) = 1 - \mu \exp(-h n),
		\end{equation}
		where $\mu$ and $h$ are the fitting parameters.
\end{itemize}

\subsection{Experimental Results}

To justify the proposed the utility functions, we use a real-world activity recognition dataset~\cite{anguita2013public} of accelerometer and gyroscope samples collected using waist-mounted smart phones. The machine learning model is to classify human activities including walking, going upstairs, going downstairs, sitting, standing, and lying down. Data samples were collected using Samsung Galaxy S II from $30$ volunteers with a sampling frequency of $50$Hz. The dataset includes $561$ calibrated features extracted using a sliding window of $2.56$sec. This window length helps detect varying-speed activities of people at different ages. For the classification algorithm, we use logistic regression and support vector machines (SVMs) with a radial basis function kernel. Figure~\ref{fig:activity_accuracy} shows the activity recognition accuracy under different data sizes. Clearly, the recognition accuracy increases as the size of training data increases. Moreover, the increase of accuracy becomes diminishing when the data size becomes larger. More importantly, the candidate utility functions can well approximate the actual accuracy results, rationalizing the concave increasing utility functions. Note that in our these two experiments while the fraction-based function fits the experimental results better than the exponential-based function, the latter results in more tractable derivation of the optimal pricing which will be presented in Section~\ref{sec:pricing}. Nonetheless, the conclustion cannot be made for general cases, and further investigation is required.

We also observe similar diminishing increase of accuracy for other machine learning algorithms such as k-nearest neighbors and linear regression. However, due to space limit, we omit them from the paper.


\section{Big Data Market Model}
\label{sec:sysmodel}


\begin{figure}[h]
\begin{center}
$\begin{array}{c} 
\epsfxsize=3.2 in \epsffile{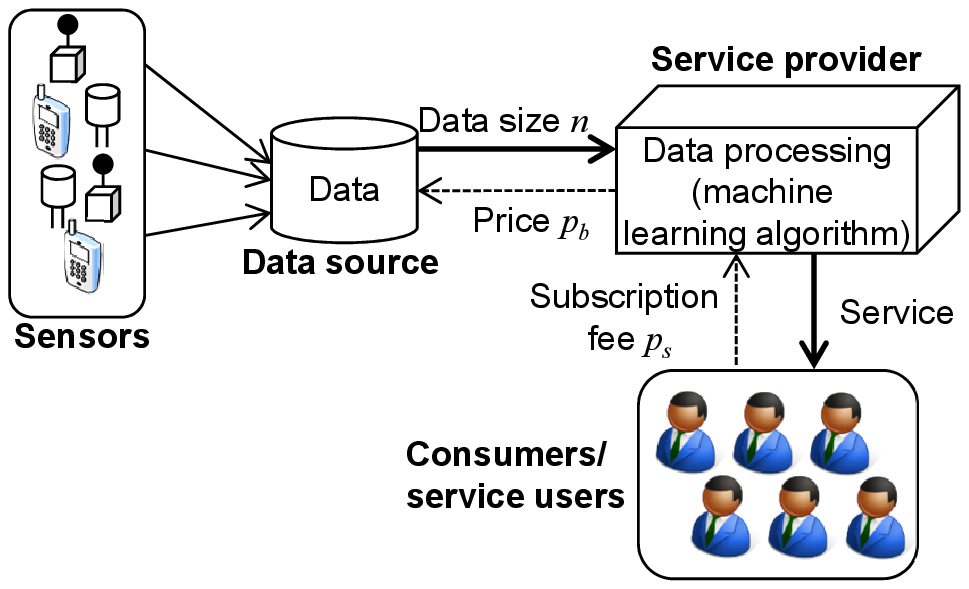} \\ [-0.2cm]
\end{array}$
\caption{Big data market model.}
\label{fig:systemmodel}
\end{center}
\end{figure}

We present and consider a typical big data market model as shown in Fig.~\ref{fig:systemmodel}. In a simplest form, the market is composed of the data source, service provider, and service consumers. The data source is responsible for gathering, e.g., from sensors, storing, maintaining, and transferring data to users. Alternatively, a service provider can obtain the data, perform data processing, e.g., machine learning and data mining algorithms, and offer services to service users or consumers. Finally, the consumers subscribe to the service offered by the provider. The consumers pay a subscription fee denoted by $p_s$ to the service provider. There are totally $M$ users, and they decide whether to subscribe to the service or not depending on their willingness-to-pay to the service and the subscription fee. The service provider buys raw data from the data source. The unit price of the raw data is denoted by $p_b$. The provider specifies the size of requested data to buy, denoted by $n$, which is defined as data demand to the data source. Here, the service provider can set the subscription fee $p_s$ and the requested data size $n$ to buy so that its profit is maximized. The data source can set the data price $p_b$ again so that its profit is maximized.

The big data market model can be applied to many big data and IoT services.
\begin{itemize}
	\item {\em Placemeter} (https://www.placemeter.com/) buys raw video data containing scenes of street and city from users. Placementer performs video analytics to extract useful information, e.g., road traffic condition and crowd, and sell the information to other businesses. Here, with reference to the big data market model, a video camera is the data source while Placemeter is the provider.
	\item {\em IoT search engine:} IoT search engine provides a generic capability for users to acquire sensing data. Sensor owners can share or sell sensing data to consumers. One example is Thingful (https://thingful.net/) that lets sensors be connected, and users can browse and obtain sensing data. The IoT search engine is able to locate, index, and make sensing data searchable. IoT sensors are the data sources, and IoT search engine is the provider.
\end{itemize}

Pricing models of DaaS are classified into four main types: (1)~request-based, (2)~volume-based, (3)~data type-based, and (4)~corporate subscription models~\cite{sarkar2015data}. In this paper, our proposed pricing scheme is the volume-based model in which the provider pays per size of data transferred from the data sources. The optimal pricing schemes for request-based, data type-based, and corporate subscription can be developed in the future work.




\section{Optimal Pricing}
\label{sec:pricing}

We first derive the profit function of the service provider in the big data market model. Then, the optimal data price of the data source is obtained for two proposed utility functions.

\subsection{Profit Function of Service Provider}

We introduce an optimal pricing scheme for the big data market model presented in Section~\ref{sec:sysmodel}. The service provider decides on the amount of data to be purchased from the data source and the subscription fee to charge the users. The service users have different willingness-to-pay for the service. If the willingness-to-pay of a user is higher than or equal to the subscription fee charged by the service provider, the user will buy (subscribe) the service. Let $w$ denote willingness-to-pay of a user to the service. The willingness-to-pay depends on the service quality. Let $w'$ denote a nominal willingness-to-pay. Then, the actual willingness-to-pay is $w = \phi( n ) w'$, where $\phi(n)$ is the data utility function of the size of input data $n$ used in the machine learning algorithms. Similarly, let $W'$ denote the maximum nominal willingness-to-pay, and thus we have $W = \phi(n) W'$ for the maximum actual willingness-to-pay. 

Given a set of users, the probability density of the willingness-to-pay is denoted by $f(w)$, where $w \in [0,W]$. Then, the expected profit of the service provider is computed as follows:
\begin{eqnarray}
	F( n, p_s )	& = &	\underset{\mathrm{revenue}} {\underbrace{  p_s M Pr ( w \geq p_s )	}}	-	\underset{\mathrm{cost}} {\underbrace{  p_b n	} } ,	\label{eq:serviceprovider_profit}	\\
				& = &	p_s M \int_{p_s}^W f(w) {\mathrm{d}} w		-	p_b n	,	
\end{eqnarray}
where the first term is the revenue gained from the subscribed users who have the willingness-to-pay higher than the subscription fee, and the second term is the cost paid to the data source. We assume that the willingness-to-pay of the users follows a uniform distribution between $[0,W']$. Thus, the profit of the service provider can be expressed as follows:
\begin{eqnarray}
	F( n, p_s )	& = &	p_s M ( W - p_s ) -	p_b n	,	\\
				& = &	p_s M ( \phi(n) W' - p_s ) -	p_b n	,
\end{eqnarray} 
where $\phi(n)$ from Section~\ref{sec:value} is adopted.

\subsection{Optimal Subscription Fee of Service Provider}

An optimization problem can be formulated to obtain an optimal subscription fee and the size of raw data to be purchased from the data source, i.e., 
\begin{equation}
	\max_{ n, p_s } F( n, p_s )		.
\label{eq:formulation}
\end{equation}

We consider two cases of different data utility functions.

\subsubsection{Case 1}

When the data utility function $\phi(n) = 1 - \frac{\kappa}{1 + g n } $ is adopted, the profit function of the service provider is expressed as follows:
\begin{eqnarray}
	F( n, p_s )	& = &	p_s M \left( 1 - \frac{\kappa}{1 + g n } - p_s \right) -	p_b n	,
\end{eqnarray} 
where without loss of generality, we set $W'=1$. By differentiating $F( n, p_s )$ with respect to $n$ and $p_s$, we have
\begin{eqnarray}
	\frac{ \partial F( n, p_s ) } { \partial n } & 	= &	\frac{ g \kappa M p_s } { ( g n + 1 )^2 } - p_b = 0	,	\\
	\frac{ \partial F( n, p_s ) } { \partial p_s } 	& = &	M p_s + M \left( p_s + \frac{ \kappa } { g n + 1  } - 1 \right)	= 0	.
\end{eqnarray}
The closed-form solutions of $n^*$ and $p^*_s$ exist. There are three roots for the solutions, i.e.,
\begin{eqnarray}
	n^*_1	& = & 	\frac{ M A_2 } { p_b }	-	\frac{ 2 M (A_2)^2 } { p_b	} - \frac{ 1 }{ g } ,	\nonumber	\\
	n^*_2	& = & 	{\mathrm{Real}} \bigg( \frac{ \kappa/48 - (A_{10})^{2/3} + A_1 } { A_9 } \nonumber	\\
			&   &	+ \frac{ g M A_{11} / 6 + A_5 + A_8 - A_3 -A4 + A_6	} { A_7 } \bigg),	\nonumber	\\
	n^*_3	& = & 	{\mathrm{Real}} \bigg( - \frac{ ( A_{10} )^{2/3} - \kappa/48 + A_1 }{ A_9 } \nonumber	\\
			&   &	\frac{ g M A_{11} /6 + A_5 + A_8 + A_3 + A_4 - A_6 } { A_7 } \bigg),
\end{eqnarray}
where ${\mathrm{Real}}(\cdot)$ returns a real value, and 
\begin{eqnarray}
	A_1		& = &	\frac{ \sqrt{3} \kappa i } { 48 }	,	\quad	A_2		= 		\frac{ 1 } { 36 (A_{10})^{1/3} } + ( A_{10} )^{1/3} + \frac{ 1 }{ 3 }	,	\nonumber \\
	A_3		& = &	\frac{ \sqrt{3} g M (A_{10})^{1/3} i }{ 216 }	,	\quad	A_4		=	\sqrt{3} g M (A_{10})^{4/3}	i ,	\nonumber \\
	A_5		& = &	\frac{ g M (A_{10})^{1/3} }{ 216 }	,			\quad	A_6		=	\frac{ \sqrt{3} g M A_{11} i } { 6 }	,	\nonumber \\
	A_7		& = &	p_b g (A_{10})^{2/3}	,	\quad A_8		=	g M (A_{10})^{4/3},	\nonumber \\
	A_9		& = &	g (A_{10})^{2/3},	\quad A_{10}		=	A_{11} + A_{12}  - \frac{ 1 }{ 216 }	,	\nonumber \\
	A_{11}		& = &	\sqrt{ ( A_{12} - 1/216 )^2 - 1/46656 },	\nonumber \\
	A_{12}		& = &	p_b \kappa / 8 g M ,	
\end{eqnarray}
and 
\begin{eqnarray}
	p_{s1}^*	& = & 	\frac{ ( 6 B_1 + 1 )^2 } { 36 B_1 }	,	\\
	p_{s2}^*	& = & 	\frac{ 1 }{ 3 } - \frac{ B_1 }{ 2 } - \frac{ 1 }{ 72 B_1 }	+ \frac{ \sqrt{3} \left( 	\frac{ 1 } { 36 B_1 } - B_1	\right) i  }  { 2 }	,	\\
	p_{s3}^*	& = & 	\frac{ 1 }{ 3 } - \frac{ B_1 }{ 2 } - \frac{ 1 }{ 72 B_1 }	- \frac{ \sqrt{3} \left( 	\frac{ 1 } { 36 B_1 } - B_1	\right) i  }  { 2 }	,	\nonumber
\end{eqnarray}
where
\begin{eqnarray}
	B_1		& = &	\sqrt[3]{	\sqrt{ \left(	\frac{ p_b \kappa } { 8 g M } - \frac{ 1 } {216 } \right)^2 - \frac{ 1 }{ 46656 } } + \frac{ p_b \kappa} { 8 g M } - \frac{ 1 } { 216 } }	.	\nonumber
\end{eqnarray}
Since the last terms of $p_{s2}^*$ and $p_{s3}^*$ are imaginary, the second and third roots of $p^*_s$ is simply $p^*_s = \frac{ 1 }{ 3 } - \frac{ B_1 }{ 2 } - \frac{ 1 }{ 72 B_1 }$.

An optimal size of raw data to be purchased by the service provider and the subscription fee are obtained from
\begin{equation}
	( n^*, p^*_s ) = \arg \max_{ (n, p_s) \in \{ n^*_1, n^*_2, n^*_3 \} \times\{ p_{s1}^*, p_{s2}^*, p_{s3}^* \} } F( n, p_s )	.
\end{equation}

We can consider the special cases that the size of raw data is fixed or the subscription fee is fixed. The former corresponds to the case that the data source has a fixed amount of raw data to supply, and the service provider optimizes only the subscription fee. In contrast, the latter corresponds to the case that the subscription fee is fixed, and the service provider optimizes only the size of raw data to be purchased. We have the following proposition.

\begin{mypro}
If $n$ is fixed, the solution $p^*_s$ of the problem in (\ref{eq:formulation}) is globally optimal. In contrast, if $p_s$ is fixed, the solution $n^*$ of the problem in (\ref{eq:formulation}) is globally optimal.
\end{mypro}
\begin{IEEEproof}
We obtain the second derivatives of $F( n, p_s )$ , i.e., 
\begin{eqnarray}
	\frac{	\partial^2 F( n, p_s ) } { \partial p^2_s }	& = &	-2 M	,	\\
	\frac{	\partial^2 F( n, p_s ) } { \partial n^2 }	& = &	- \frac{ 2 g^2 \kappa M p_s } { (n g + 1 )^3 } 	,
\end{eqnarray}
which are non-positive. Therefore, the solutions of the special cases are globally optimal. 
\end{IEEEproof}

\subsubsection{Case 2}

When the data utility function $1 - \mu \exp(-h d)$ is adopted, the profit function of the service provider is expressed as follows:
\begin{eqnarray}
	F( n, p_s )	& = &	p_s M \left( 1 - \mu \exp(-h n) - p_s \right) -	p_b n	.
\end{eqnarray} 
By differentiating $F( n, p_s )$ with respect to $n$ and $p_s$, we have
\begin{eqnarray}
	\frac{ \partial F( n, p_s ) } { \partial n } & 	= &	h \mu M p_s \exp ( - h n ) - p_b = 0	,	\\
	\frac{ \partial F( n, p_s ) } { \partial p_s } 	& = &	M ( 2 p_s + \mu \exp ( -h n ) - 1 )	= 0	.
\end{eqnarray}
The closed-form solutions of $n^*$ and $p^*_s$ exist. There are two roots for the solutions, i.e.,
\begin{eqnarray}
	n^*	& = & 	\frac{ \ln \left( \frac{ h \mu M}{ 4 p_b }  \pm \frac{ h \mu M \sqrt{ \frac{1}{4} - \frac{2 p_b}{ h M } } }{ 2 p_b} \right) }{ h }		,
\end{eqnarray}
and 
\begin{eqnarray}
	p_s^*	& = & 	\frac{1}{4} \pm \frac{ \sqrt{ 1- \frac{ 8 p_b }{ h M } } }{4} 	.
\end{eqnarray}

Again, we consider the special cases that $n$ is fixed or $p_s$ is fixed. The second derivatives of $F( n, p_s )$ are
\begin{eqnarray}
	\frac{	\partial^2 F( n, p_s ) } { \partial p^2_s }	& = &	-2 M	,	\\
	\frac{	\partial^2 F( n, p_s ) } { \partial n^2 }	& = &	 -h^2 \mu M p_s \exp ( -h n )	,
\end{eqnarray}
which are non-positive. Therefore, the solutions of the special cases are globally optimal.

\subsection{Optimal Pricing Scheme of Data Source}

The data source can adjust data price $p_b$ so that its profit is maximized. We can formulate a Stackelberg game to model this situation. 
\begin{itemize}
	\item The leader player is the data source, the strategy of which is data price $p_b$. The payoff is the profit defined as $P ( p_b ) = p_b n(p_b)$.
	Here, the size of data to be purchased $n( p_b)$ is defined as a function of $p_b$. 
	\item The follower player is the service provider, the strategy of which is the amount of data to be purchased and the subscription fee. The payoff is profit $F( n, p_s )$ defined as (\ref{eq:serviceprovider_profit}).
\end{itemize}
The Stackelberg equilibrium is defined as follows:
\begin{eqnarray}
	p_b^*			& = &	\arg \max_{ p_b } 	p_b n(p_b)	,	\\
	\mbox{s.t.}		&	&	n (p_b) = 	\arg \max_{ n, p_s } F( n, p_s )	.
\end{eqnarray}


\section{Numerical examples}
\label{sec:performance}

We consider the data utility function $\phi(n) = 1 - \frac{\kappa}{1 + g n }$ to obtain the representative numerical results of the optimal pricing scheme. We use the following setting: the number of users is 500. We adopt the fitted parameters as shown in Fig.~\ref{fig:activity_accuracy}(a). Note that similar results can be obtained for the other utility function with different classification machine learning algorithms.



\begin{figure}[h]
\begin{center}
$\begin{array}{c} 
\epsfxsize=3.4 in \epsffile{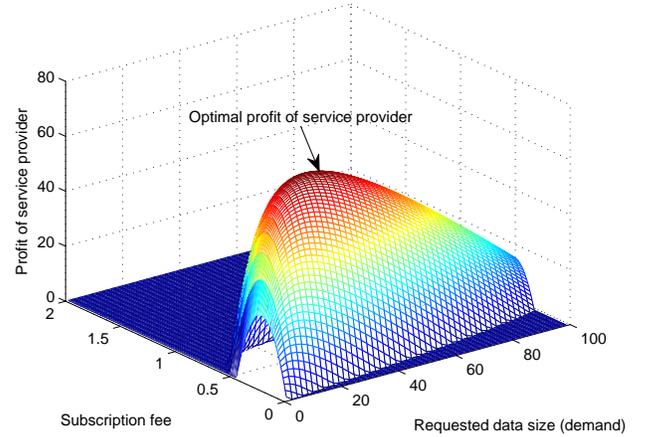} \\ [-0.2cm]
\end{array}$
\caption{Profit of service provider under varied requested data sizes and subscription fee strategies.}
\label{fig:optimal_profit}
\end{center}
\end{figure}

Figure~\ref{fig:optimal_profit} shows the profit of the service provider when under different requested data sizes and subscription fee strategies. When the requested data size is small, the service quality is poor and the utility is low, and thus only few users will subscribe the service from the provider. However, if the requested data size is large, the service quality becomes better, but the provider has to pay more for the data. Likewise, if the subscription fee is high, only few users will subscribe the service, resulting in a small revenue. By contrast, if the fee is low, the profit will be adversely affected. Clearly, there is a maximum profit which can be achieved when the optimal requested data size and subscription fee are applied. These optimal solutions can be obtained from the presented closed-form expressions.


\begin{figure}[h]
\begin{center}
$\begin{array}{c} 
\epsfxsize=3.4 in \epsffile{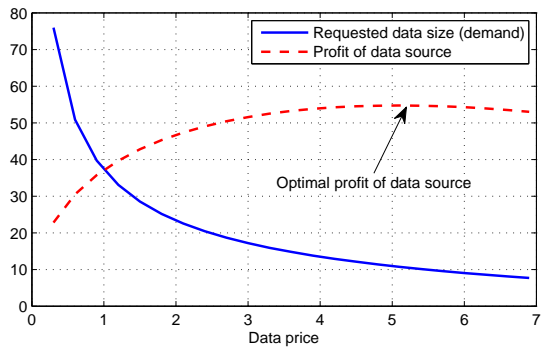} \\ [-0.2cm]
\end{array}$
\caption{Optimal requested data size and profit of data source under varied data price.}
\label{fig:vary_dataprice}
\end{center}
\end{figure}

Figure~\ref{fig:vary_dataprice} shows the optimal requested data size, which represents demand from the service provider to the data source. Clearly, as the data price increases, the demand decreases. Given this demand function, the figure also shows the profit of the data source. Again, there is an optimal profit that can be achieved when an optimal data price is applied. This optimal data price as well as the optimal requested data size and subscription fee form the Stackelberg equilibrium points for the data source, i.e., a leader, and the service provider, i.e., a follower, respectively.

\begin{figure}[h]
\begin{center}
$\begin{array}{c} 
\epsfxsize=3.4 in \epsffile{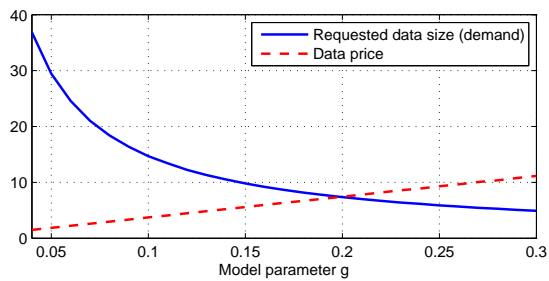} \\ [-0.2cm]
\end{array}$
\caption{Optimal requested data size and profit of data source under varied model parameter $g$.}
\label{fig:vary_modelparameter}
\end{center}
\end{figure}

Next, we vary some model parameters of the data utility function. By increasing $g$, the data-service quality improves faster given the same size of data used in the machine learning algorithm. Thus, from Fig.~\ref{fig:vary_modelparameter}, the requested data size or demand decreases. Here, to maximize the profit, the data source can raise the data price.



\section{Summary}
\label{sec:summary}

In this paper, we have studied economic issues of big data and IoT. Specifically, we have proposed utility functions of data when the data is used in big data analytics. Through the case study of classification-based machine learning algorithms, the suitability of the functions has been demonstrated. Next, we have introduced a big data market model, which is composed of a data source, a service provider, and users. Based on the data utility functions, we have developed an optimal pricing scheme that allows the service provider to determine the amount of data to be acquired to provide services to the users. Additionally, we have shown that the Stackelberg game can model the strategy of the data source to achieve the maximum profit.


\section*{Acknowledgements}
This work was supported in part by the National Research Foundation of Korea (NRF) grant funded by the Korean government (MSIP) (2014R1A5A1011478) and Singapore MOE Tier 1 (RG18/13 and RG33/12) and MOE Tier 2 (MOE2014-T2-2-015 ARC 4/15).


\end{document}